\documentclass[journal]{IEEEtran}
\usepackage{graphicx,amssymb,amsmath}%ͼƬ֧³Ö
\usepackage{multicol}%ÕýÎÄË«À¸
\usepackage[noadjust]{cite}%²Î¿¼ÎÄÏ×
\usepackage{setspace}%µ÷ÕûÐоà
\usepackage{stfloats}
\usepackage{midfloat}
\usepackage[normal]{threeparttable}
\usepackage{flushend,cuted}
\usepackage{cuted}
\usepackage{cases,subeqnarray}
\usepackage{bm,multirow,bigstrut}
\usepackage{textcomp}
\usepackage{latexsym,bm}
\usepackage{booktabs,changebar}
\usepackage{xcolor}
\usepackage{mathtools}

\begin{document}
\title{Unified Performance Analysis of Mixed Radio Frequency/Free-Space Optical Dual-Hop Transmission Systems}

\author{Jiayi~Zhang,~\IEEEmembership{Member,~IEEE},
        Linglong~Dai,~\IEEEmembership{Senior Member,~IEEE},
        Yu~Zhang,~\IEEEmembership{Senior Member,~IEEE},
        and~Zhaocheng~Wang,~\IEEEmembership{Senior Member,~IEEE}
 
\thanks{
This work was supported by the National Key Basic Research Program of China (No. 2013CB329203), National Natural Science Foundation of China (Grant No. 61201185), National High Technology Research and Development Program of China (Grant No. 2014AA01A704), and China Postdoctoral Science Foundation (No. 2014M560081).}%
\thanks{J. Zhang, L. Dai (corresponding author), Y. Zhang and Z. Wang are with Department of Electronic Engineering as well as Tsinghua National Laboratory for Information Science and Technology (TNList), Tsinghua University, Beijing 100084, P. R. China (e-mails: \{jiayizhang, daill, zhang-yu, zcwang\}@tsinghua.edu.cn).}
\thanks{Copyright (c) 2015 IEEE. Personal use of this material is permitted.  However, permission to use this material for any other purposes must be obtained from the IEEE by sending a request to pubs-permissions@ieee.org.}}

\maketitle

\begin{abstract}
The mixed radio frequency (RF)/free-space optical (FSO) relaying is a promising technology for coverage improvement, while there lacks unified expressions to describe its performance. In this paper, a unified performance analysis framework of a dual-hop relay system over asymmetric RF/FSO links is presented. More specifically, we consider the RF link follows generalized $\kappa$-$\mu$ or $\eta$-$\mu$ distributions, while the FSO link experiences the gamma-gamma distribution, respectively. Novel analytical expressions of the probability density function and cumulative distribution function are derived. We then capitalize on these results to provide new exact analytical expressions of the outage probability and bit error rate (BER). Furthermore, the outage probability for high signal-to-noise ratios and the BER for different modulation schemes are deduced to provide useful insights into the impact of system and channel parameters of the overall system performance. These accurate expressions are general, since they correspond to generalized fading in the RF link and account for pointing errors, atmospheric turbulence and different modulation schemes in the FSO link. The links between derived results and previous results are presented. Finally, numerical and Monte-Carlo simulation results are provided to demonstrate the validity of the proposed unified expressions.
\end{abstract}

\begin{IEEEkeywords}
Free--space optical communications, outage probability, bit error rate, atmospheric turbulence.
\end{IEEEkeywords}

%%%%%%%%%%%%%%%%%%%%%%%%%%  body  %%%%%%%%%%%%%%%%%%%%%%%%%%
\section{Introduction}
Due to the merits of free license, low cost and high bandwidth, free-space optical (FSO) communication is becoming one of the promising technologies for indoor and outdoor wireless applications. Some typical applications of FSO systems include ``last mile" access, indoor positioning, disaster recovery, military applications, underwater system, device-to-device communications, and video transmission, etc. \cite{simpson2012smart,ghassemlooy2012optical,chatzidiamantis2011inverse}.
On the other hand, dual-hop relaying has been widely adopted in the context of radio frequency (RF) communication systems because of the extended coverage area and enhanced receive signal induced by spatial diversity. In traditional RF/RF dual-hop communication systems, the scarcity of licensed spectrum in conjunction with fast growing demand of high data rate is one of the most crucial limitations. To this end, the FSO communication has been considered to be used in the relaying as a complement to RF counterpart due to its desirable features: First, the RF and FSO links operate on different frequency bands to avoid significant interference in RF/RF relay systems; Second, the FSO link can achieve maximum system capacity by aggregating multiple users' information.

Therefore, the mixed RF/FSO dual-hop relay system has been recently proposed \cite{Lee2011performance,Ansari2013impact,samimi2013end,Zedini2015performance,soleimani2014generalized} and refers to the case when RF transmission is used at one hop and FSO transmission at the other. Such topology is quite different from hybrid RF/FSO system, which uses parallel RF and FSO links for the same path. In \cite{Lee2011performance}, a performance analysis of a dual-hop relay system composed of mixed RF/FSO links was first conducted. The impact of pointing errors on such topologies has been analyzed in \cite{Ansari2013impact}. The authors in \cite{samimi2013end} derived an exact closed-form expression of the end-to-end outage probability of the mixed RF/FSO system. Moreover, the performance of dual-hop systems with Nakagami-$m$ and gamma-gamma ($\Gamma\Gamma$) or double GG \cite{Kashani2013novel} fading channels was investigated in \cite{Zedini2015performance,soleimani2014generalized}.

Although the performance of mixed RF/FSO relaying systems over fading channels has been extensively evaluated in terms of outage probability and error rate, most of existing pioneering studies simply assume the Rayleigh or Nakagami-$m$ fading channels for the RF link. While such assumption extensively simplifies some mathematical manipulations, these distributions fall short of capturing the actual fading statistics of the RF link, because their tail does not seem to yield a good fit to experimental data \cite{parsons2000mobile}. Motivated by this fact, in this work we examine the performance of a dual-hop RF/FSO transmission system experiencing more general $\kappa$-$\mu$/$\Gamma\Gamma$ and $\eta$-$\mu$/$\Gamma\Gamma$ fading conditions. The $\eta$-$\mu$ and $\kappa$-$\mu$ fading were proposed in \cite{yacoub2007kappa} and has been verified to provide an excellent fit to experimental data. Moreover, it provides a unified framework for existing RF channel models, since it regards the Rayleigh ($\eta \rightarrow 0, \kappa \rightarrow 0, \mu=1$), Rician ($\mu=1$) and Nakagami-$m$ ($\eta \rightarrow 0,\kappa \rightarrow 0,\mu=m$) distributions as special cases. For the FSO link, we use the $\Gamma\Gamma$ distribution to describe the weak or strong atmospheric turbulence fading, since the $\Gamma \Gamma$ distribution has simpler statistical characteristics than the double GG distribution, and has been proved to provide a good agreement between the theoretical result and the corresponding experimental data \cite{andrews2005laser,wang2010moment,al2001mathematical}. Moreover, the pointing errors, which is caused by an unavoidable alignment between transmitter and receiver due to thermal expansion, dynamic wind loads, and weak earthquakes, is one of the major issues leading to a severe performance degradation of the FSO link \cite{jurado2012impact,garcia2012bit}. Therefore, we study the effect of pointing errors on the performance of dual-hop RF/FSO systems.

The contributions of this paper are summarized as follows:

\begin{itemize}
\item Novel analytical expressions of the probability density function (PDF) and the cumulative distribution function (CDF) for the mixed RF/FSO dual-hop relaying system are derived in terms of Meijer's $G$-function, which can be easily evaluated and efficiently programmed in most standard software packages (e.g. MAPLE, MATHEMATICA). We further demonstrate that previous results in \cite{Lee2011performance,Zedini2015performance,Ansari2013impact} for the Rayleigh and Nakagami-$m$ RF fading channels can be considered as special cases of our results.

\item We derive the generalized analytical expressions of the end-to-end outage probability and average bit error rate (BER) for the mixed RF/FSO dual-hop system. These exact formulations are more general, since they correspond to $\kappa$-$\mu$ and $\eta$-$\mu$ fading in the RF link, and account for pointing errors, atmospheric turbulence and different modulation schemes in the FSO link.

\item The asymptotic closed-form outage probability in the high signal-to-noise ratio (SNR) regime is assessed, which can provide useful insights into the impact of system and fading parameters on the overall system performance. From an engineering perspective, we particularly investigate the effect of pointing errors in the FSO link on the BER performance, over different modulation schemes and/or various channel fading conditions.

\item Although corresponding expressions for the case of $\kappa$-$\mu$/$\Gamma\Gamma$ fading presented herein are given in terms of infinite series, we only need less than ten terms to get a satisfactory accuracy (e.g., smaller than $10^{-6}$) for all considered cases.

\end{itemize}

The remainder of the paper is structured as follows: In Section \ref{se:system_model} the mixed RF/FSO system model is presented, and the statistical characteristics of generalized fading channels are described in Section \ref{se:statistical_characteristics}. In Section \ref{se:performance_analysis}, we derive novel and analytical expressions of the outage probability and average BER for the mixed RF/FSO dual-hop system. Numerical and Monte-Carlo simulation results are provided in Section \ref{se:numerical_results}, and Section \ref{se:conclusion} concludes with a summary of the main results.

\section{Mixed RF-FSO System Model}\label{se:system_model}

\begin{figure}[t]
\centering
\includegraphics[scale=0.65]{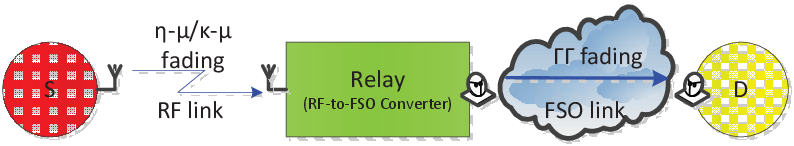}
\caption{The mixed RF/FSO dual-hop relay system with S, R and D denoting the source node, relay node and destination node, respectively. Moreover, the first hop (RF link) experiences a $\kappa$-$\mu$ or $\eta$-$\mu$ fading, and the second hop (FSO link) follows a $\Gamma\Gamma$ fading.
\label{fig:RF_FSO_model}}
\end{figure}

As shown in Fig. \ref{fig:RF_FSO_model}, we consider a dual-hop relay system with the source node S, the relay node R and the destination node D. The source node S cannot communicate directly with the destination node D because of the long distance or the block between them. The RF link between S and R experiences a $\kappa$-$\mu$ and $\eta$-$\mu$ fading, since the $\kappa$-$\mu$ and $\eta$-$\mu$ fading is more general and includes the Rayleigh, Rician and Nakagami-$m$ fading as special cases \cite{yacoub2007kappa}. While the FSO link between R and D follows a $\Gamma\Gamma$ distribution, which is widely used in FSO communications \cite{ghassemlooy2012optical}. Similar to the related literature \cite{bayaki2009performance,chatzidiamantis2011distribution,aggarwal2014dual}, we consider the high-energy FSO system with the commonly used on-off keying modulation. In this case, the FSO system performance is limited by background radiation and thermal noise, which can be modeled as the additive white Gaussian noise (AWGN) with the variance ${N_0}/2$ as a good approximation of the Poisson photon-counting detection model \cite{navidpour2007ber}. Furthermore, we consider the typical amplify-and-forward (AF) relaying scheme in the relay node R, which simply amplifies its received signal by a fixed gain $G$ and then forward it to the destination node D. Following the similar analysis in \cite{peppas2013capacity}, the end-to-end received instantaneous electrical SNR ${\gamma _\text{eq}}$ at the destination node D is given by
\begin{align}
{\gamma _\text{eq}} = \frac{{{\gamma _1}{\gamma _2}}}{{c + {\gamma _2}}},
\end{align}
where $c\triangleq G^2/{N_0}$, $\gamma _1$ denotes the SNR of the RF hop (i.e. S-R link) and $\gamma _2$ represents the electrical SNR of the FSO hop (i.e. R-D link).

\section{Statistical Characteristics}\label{se:statistical_characteristics}
In this section, we first deduce the novel analytical CDF expression for the mixed RF/FSO dual-hop system. Then, the PDF expression is derived by differentiating the CDF expression. These expressions are useful for following analysis.

\subsection{Cumulative Distribution Function}
The CDF of the end-to-end instantaneous received electrical SNR ${\gamma _\text{eq}}$ can be expressed as
\begin{align}\label{eq:cdf_total_snr}
{F_{{\gamma _\text{eq}}}}\left( x \right) &= \int_0^\infty  {\Pr \left[ {\left. {{\gamma _\text{eq}} < x} \right|{\gamma _2}} \right]} {f_{{\gamma _2}}}\left( {{\gamma _2}} \right)d{\gamma _2} \notag \\
&= \int_0^\infty  {{F_{{\gamma _1}}}\left( {\frac{{c + {\gamma _2}}}{{{\gamma _2}}}x} \right)} {f_{{\gamma _2}}}\left( {{\gamma _2}} \right)d{\gamma _2}.
\end{align}
Note that \eqref{eq:cdf_total_snr} involves the PDF of the SNR at the relay node R, and the CDF of the electrical SNR at the destination node D. The PDF of $\Gamma\Gamma$ fading channels with pointing errors is given by \cite{Sandalidis2009optical,gappmair2011further}
\begin{align}\label{eq:pdf_gammagamma}
{f_{\Gamma \Gamma }}\left( {\gamma _2}\right) = \frac{{{\xi ^2}}}{{t\Gamma \left( a \right)\Gamma \left( b \right){\gamma _2}}}G_{1,3}^{3,0}\left[ {\left. {dab{{\left( {\frac{{{{\gamma _2}}}}{{{\kappa _t}}}} \right)}^{1/t}}} \right|\begin{array}{*{20}{c}}
{{\xi ^2} + 1}\\
{{\xi ^2},a,b}
\end{array}} \right],
\end{align}
where $G(\cdot)$ is the Meijers's $G$-function \cite[Eq. (9.301)]{gradshtein2000table}, $\Gamma(\cdot)$ is the gamma function \cite[Eq. (8.310)]{gradshtein2000table}, $t$ accounts for the widely used detection technique type (i.e. $t=1$ represents heterodyne detection, and $t=2$ is intensity modulation with direct detection (IM/DD)), $d={\xi ^2}/({\xi ^2}+1)$ with $\xi $ being the ratio between the equivalent beam radius and the pointing errors displacement standard deviation at the destination node D, and ${\kappa _t}$ denotes the average SNR as \cite{Ansari2013impact,Sandalidis2009optical}
%\begin{align}\label{eq:kappa_t}
%{\kappa_1} &= {\text{E}}\left( {{\gamma _2}} \right) = {{\bar \gamma }_2}\\
%{\kappa_2} &= \frac{{{\bar \gamma }_2}ab{\xi ^2}\left( {{\xi ^2} + 2} \right)}{  {\left( {a + 1} \right)\left( {b + 1} \right){{\left( {{\xi ^2} + 1} \right)}^2}} }
%\end{align}
\begin{equation}\label{eq:kappa_t}
{\kappa_t} \triangleq\left \{
\begin{aligned}
 &{\text{E}}\left( {{\gamma _2}} \right) = {{\bar \gamma }_2},   &t=1\\
 & \frac{{{\bar \gamma }_2}ab{\xi ^2}\left( {{\xi ^2} + 2} \right)}{  {\left( {a + 1} \right)\left( {b + 1} \right){{\left( {{\xi ^2} + 1} \right)}^2}} },   &t=2
\end{aligned}
\right.
\end{equation}
Note that larger effect of pointing errors means smaller values of $\xi$. Moreover, the two effective numbers $a$ and $b$ are related to the atmospheric conditions and can be expressed as \cite{bayaki2009performance}
\begin{align}
a  &= {\left( {\exp \left[ {\frac{{0.49\sigma _2^2}}{{{{\left( {1 + 0.18{d^2} + 0.56\sigma _2^{12/5}} \right)}^{7/6}}}}} \right] - 1} \right)^{ - 1}},\label{eq:beta_OC} \\
b &= {\left( {\exp \left[ {\frac{{0.51\sigma _2^2{{\left( {1 + 0.69\sigma _2^{12/5}} \right)}^{ - 5/6}}}}{{{{\left( {1 + 0.9{d^2} + 0.62{d^2}\sigma _2^{12/5}} \right)}^{5/6}}}}} \right] - 1} \right)^{ - 1}},\label{eq:m_OC}
\end{align}
where $\sigma _2^2 = 0.492C_n^2{\hat{k}^{7/6}}{L^{11/6}}$ is the Rytov variance, $d = \sqrt {\hat{k}{D^2}/4L} $ with $L$ being the distance between transmitter and receiver, $\hat{k} \triangleq 2\pi/\lambda$ denotes the optical wavenumber with $\lambda$ being the operational wavelength, $D$ is the aperture diameter of the receiver, and $C_n^2$ is the altitude-dependent index of the refractive structure parameter determining the turbulence strength. Furthermore, we assume $C_n^2$ remains constant for a relatively long transmit bits interval, and it varies from $1 \times 10^{-15} {\text{m}}^{-2/3}$ to $3\times10^{-14} {\text{m}}^{-2/3}$ for weak to strong turbulence cases \cite{navidpour2007ber}.

Then we introduce the statistical characteristics of the generalized fading for the RF hop. There are two formats for the $\eta$-$\mu$ distribution, namely, Format 1 and Format 2. According to \cite{yacoub2007kappa}, Format 1 can be converted into Format 2 through a bilinear transformation. Without loss of generality, we focus on Format 1 in the following. Due to its mathematical briefness and significant features \cite{peppas2013dual}, the common case of integer values of $\mu$ is considered. The CDF of the $\eta$-$\mu$ distribution for integer values of $\mu$ is given by \cite[Eq. (3)]{peppas2013dual}
\begin{align}\label{eq:cdf_etamu}
{F_{\eta-\mu }}\left( {\gamma _1} \right) &= 1 - \frac{1}{{\Gamma \left( \mu  \right)}}{\left( {\frac{h}{H}} \right)^\mu }\sum\limits_{n = 1}^2 \notag \\
&\times {\sum\limits_{k = 0}^{\mu  - 1}{\sum\limits_{l = 0}^{\mu  - k - 1} {\frac{{A_n^l{a_{n,k}}}}{{l!}}} } } {{\gamma _1}^l}\exp \left( { - {A_n}{\gamma _1}} \right),
\end{align}
where $h \triangleq \left( {2 + {\eta ^{ - 1}} + \eta } \right)/4$, $H \triangleq \left( {{\eta ^{ - 1}} - \eta } \right)/4$, ${A_1} \triangleq 2\mu \left( {h - H} \right)/\bar {\gamma_1} $, ${A_2} \triangleq 2\mu \left( {h + H} \right)/\bar {\gamma_1}$, ${a_{1,k}} \triangleq \frac{{{{\left( { - 1} \right)}^k}\Gamma \left( {\mu  + k} \right){H^{ - k}}}}{{{2^{\mu  + k}}k!{{\left( {h - H} \right)}^{\mu  - k}}}}$\footnote{Note that there is a typo for $a_{1,k}$ in \cite[Eq. (3)]{peppas2013dual}, where  ${{\left( {H - h} \right)}^{\mu  - k}}$ in the denominator of $a_{1,k}$ should be ${{\left( {h - H} \right)}^{\mu  - k}}$.}, ${a_{2,k}} \triangleq \frac{{{{\left( { - 1} \right)}^\mu }\Gamma \left( {\mu  + k} \right){H^{ - k}}}}{{{2^{\mu  + k}}k!{{\left( {H + h} \right)}^{\mu  - k}}}}$, and $\bar {\gamma_1}$ denotes the average SNR of the RF hop.

Moreover, the $\kappa$-$\mu$ SNR PDF is given by \cite[Eq. (2)]{yacoub2007kappa}
\begin{align}\label{eq:pdf_kappamu}
{f_{\kappa  - \mu }}\left( \gamma_1  \right) &= \frac{{\mu {{\left( {1 + \kappa } \right)}^{\frac{{\mu  + 1}}{2}}}}}{{{\kappa ^{\frac{{\mu  - 1}}{2}}}\exp \left( {\kappa \mu } \right){{\bar \gamma }_1}^{\frac{{\mu  + 1}}{2}}}}{\gamma_1 ^{\frac{{\mu  - 1}}{2}}}\exp \left( { - \frac{{\mu \left( {1 + \kappa } \right)\gamma_1 }}{{{{\bar \gamma }_1}}}} \right)\notag \\
&\times {I_{\mu  - 1}}\left( {2\mu \sqrt {\frac{{\kappa \left( {1 + \kappa } \right)\gamma }}{{{{\bar \gamma }_1}}}} } \right),
\end{align}
where $I_v(\cdot)$ is the modified Bessel function of the first kind with order $v$.
After using the infinite series representation of $I_v(\cdot)$ \cite[Eq. (8.445)]{gradshtein2000table} and simple algebraic operation, the CDF of the $\kappa$-$\mu$ distribution can be written as
\begin{align}\label{eq:cdf_kappamu}
{F_{\kappa  - \mu }}\left( \gamma_1  \right) ={e^{ - k\mu }}\sum\limits_{i = 0}^\infty  {\frac{{{{\left( {\kappa \mu } \right)}^i}}}{{i!}}} \left( {1 - \frac{{\Gamma \left( {\mu  + i,A\gamma_1 } \right)}}{{\Gamma \left( {\mu  + i} \right)}}} \right),
\end{align}
where $A \triangleq \frac{{\mu \left( {1 + \kappa } \right)}}{{{{\bar \gamma }_1}}}$ and $\Gamma(\cdot,\cdot)$ is the incomplete gamma function \cite[Eq. (8.350.2)]{gradshtein2000table}.

For the case of $\eta$-$\mu$/$\Gamma\Gamma$ fading, we substitute \eqref{eq:pdf_gammagamma} and \eqref{eq:cdf_etamu} into \eqref{eq:cdf_total_snr} and derive
\begin{align}\label{eq:cdf_total_snr1}
{F_{{\gamma _{\text{eq}}}}}\left( {\gamma} \right) &= 1 - \frac{{{\xi ^2}}}{{t\Gamma \left( \mu  \right)\Gamma \left( a \right)\Gamma \left( b \right)}}{\left( {\frac{h}{H}} \right)^\mu }\notag \\
 &\times\sum\limits_{n = 1}^2 {\sum\limits_{k = 0}^{\mu  - 1} {\sum\limits_{l = 0}^{\mu  - k - 1} {\frac{{A_n^l{a_{n,k}}{{\gamma} ^l}\exp \left( { - {A_n}{\gamma} } \right)}}{{l!}}} } }  \notag \\
 &\times \int_0^\infty  {{{\left( {1 + \frac{c}{{{\gamma _2}}}} \right)}^l}\exp \left( { - \frac{{{A_n}c{\gamma} }}{{{\gamma _2}}}} \right)} \frac{1}{{{\gamma _2}}}\notag \\
 &\times G_{1,3}^{3,0}\left[ {\left. {dab\left( {\frac{{{\gamma _2}}}{{{\kappa _t}}}} \right)^{1/t}} \right|\begin{array}{*{20}{c}}
{{\xi ^2} + 1}\\
{{\xi ^2},a,b}
\end{array}} \right]d{\gamma _2}.
\end{align}
Utilizing the binomial expansion \cite[Eq. (1.111)]{gradshtein2000table} and the integral identity as derived in \eqref{eq:integral_result}, the corresponding analytical CDF of $\gamma_{\text{eq}}$ can be derived as
\begin{align}\label{eq:cdf_total_snr_result}
{F_{{\gamma _{\text{eq}}}}^{{\eta-\mu}}}\left( \gamma  \right) &= 1 - \frac{{{\xi ^2}{t^{a + b - 2}}}}{{\Gamma \left( \mu  \right)\Gamma \left( a \right)\Gamma \left( b \right){{\left( {2\pi } \right)}^{t - 1}}}}{\left( {\frac{h}{H}} \right)^\mu }\notag \\
 &\times \sum\limits_{n = 1}^2 {\sum\limits_{k = 0}^{\mu  - 1} {\sum\limits_{l = 0}^{\mu  - k - 1} {\sum\limits_{j = 0}^l {\frac{{{a_{n,k}}{{\left( {{A_n}\gamma } \right)}^{l - j}}\exp \left( { - {A_n}\gamma } \right)}}{{j!\left( {l - j} \right)!}}} } } } \notag \\
&\times G_{t,3t + 1}^{3t + 1,0}\left[ {\left. {\frac{{{{\left( {dab} \right)}^t}{A_n}c}}{{{\kappa _t}{t^{2t}}}}\gamma } \right|\begin{array}{*{20}{c}}
{\omega}\\
{\tau}
\end{array}} \right],
\end{align}
where $\omega \triangleq \{\frac{{{\xi ^2} + 1}}{t} ,\frac{{{\xi ^2} + t}}{t}\}$ and $\tau \triangleq \{\frac{{{\xi ^2}}}{t} ,\frac{{{\xi ^2} + t - 1}}{t},\frac{a}{t} ,\frac{{a + t - 1}}{t},\frac{b}{t} ,\frac{{b + t - 1}}{t},j\}$.

For the case of $\kappa$-$\mu$/$\Gamma\Gamma$ fading, we utilize the finite series representation of $\Gamma(\cdot,\cdot)$ \cite[Eq. (8.352.4)]{gradshtein2000table} and the similar method aforementioned. Then the corresponding analytical CDF of $\gamma_{\text{eq}}$ can be derived as
\begin{align}\label{eq:cdf_total_snr_result_kappa_mu}
&{F_{{\gamma _{\text{eq}}}}^{{\kappa-\mu}}}\left( \gamma  \right)= \sum\limits_{i = 0}^\infty  {\frac{{{{\left( {\kappa \mu } \right)}^i}}}{{i!{e^{k\mu }}}}}  - \frac{{{\xi ^2}{t^{a + b - 2}}}}{{{{\left( {2\pi } \right)}^{t - 1}}\Gamma \left( a \right)\Gamma \left( b \right){e^{A\gamma  + \kappa \mu }}}}\notag \\
&\times \sum\limits_{i = 0}^\infty  {\sum\limits_{l = 0}^{\mu  + i - 1} {\sum\limits_{j = 0}^l {\frac{{{{\left( {A\gamma } \right)}^{l - j}}{{\left( {\kappa \mu } \right)}^i}}}{{i!j!\left( {l - j} \right)!}}} } G_{t,3t + 1}^{3t + 1,0}\left[ {\left. {\frac{{{{\left( {dab} \right)}^t}Ac}}{{{\kappa _t}{t^{2t}}}}\gamma } \right|\begin{array}{*{20}{c}}
\omega \\
\tau
\end{array}} \right]} .
\end{align}
Although the CDF expression of $\kappa$-$\mu$/$\Gamma\Gamma$ fading channels is given in the form of infinite series, we can truncated \eqref{eq:cdf_total_snr_result_kappa_mu} suitably (no more than ten terms) so as to achieve a satisfactory accuracy (e.g., smaller than ${10^{-6}}$) as seen in Fig. \ref{fig:OP_kappamu}.

It is worthy to point out that for $\eta\rightarrow 0$, $\kappa \rightarrow 0$ and $\mu=m$ (or $\eta\rightarrow 1$ and $\mu=m/2$), \eqref{eq:cdf_total_snr_result} and \eqref{eq:cdf_total_snr_result_kappa_mu} reduce to the CDF of the mixed Nakagami-$m$/$\Gamma\Gamma$ dual-hop relaying systems as
\begin{align}
& {F_{{\gamma _{\text{eq}}}}^ {\text{Nakagami}}}\left( \gamma  \right) = 1 - \frac{{{t^{a + b - 2}}{\xi ^2}}}{{{{\left( {2\pi } \right)}^{t - 1}}\Gamma \left( a \right)\Gamma \left( b \right)}}\exp \left( { - \frac{m}{{{{\bar \gamma }_1}}}\gamma } \right)\notag \\
 &\times \sum\limits_{l = 0}^{m \!-\! 1} {\sum\limits_{j = 0}^l {\frac{1}{{j!\left( {l \!-\! j} \right)!}}} } {\left( {\frac{m\gamma}{{{{\bar \gamma }_1}}} } \right)^{l \!-\! j}} G_{t,3t \!+\! 1}^{3t \!+\! 1,0}\left[ {\left. {\frac{{{{\left( {dab} \right)}^t}c}}{{{\kappa _t}{t^{2t}}}}\frac{m}{{{{\bar \gamma }_1}}}\gamma } \right|\begin{array}{*{20}{c}}
{\omega}\\
{\tau}
\end{array}} \right], \notag
\end{align}
which is in agreement with \cite[Eq. (8)]{Zedini2015performance}. Furthermore, for dual-hop relaying systems over mixed Rayleigh/$\Gamma\Gamma$ fading channels, e.g. $\eta\rightarrow 0$, $\kappa\rightarrow 0$ and $\mu=1$, our result presented in \eqref{eq:cdf_total_snr_result} and \eqref{eq:cdf_total_snr_result_kappa_mu} can be shown to agree with \cite[Eq. (2)]{Ansari2013impact}.

\subsection{Probability Density Function}
The PDF expression can be obtained by differentiating \eqref{eq:cdf_total_snr_result} and \eqref{eq:cdf_total_snr_result_kappa_mu} with respect to $\gamma$. With the help of \cite[Eq. (07.34.20.0001.01)]{Wolfram2011function} and \cite[Eq. (9.31.5)]{gradshtein2000table}, we derive the analytical PDF expressions for two cases as
\begin{align}\label{eq:pdf_total_snr_result}
&{f_{{\gamma _{\text{eq}}}}^{{\eta-\mu}}}\left( \gamma  \right)= \frac{{{\xi ^2}{t^{a + b - 2}}}}{{\Gamma \left( \mu  \right)\Gamma \left( a \right)\Gamma \left( b \right){{\left( {2\pi } \right)}^{t - 1}}}}{\left( {\frac{h}{H}} \right)^\mu }\notag \\
 &\times \sum\limits_{n = 1}^2 {\sum\limits_{k = 0}^{\mu  - 1} {\sum\limits_{l = 0}^{\mu  - k - 1} {\sum\limits_{j = 0}^l {\frac{{{a_{n,k}}{{\left( {{A_n}\gamma } \right)}^{l - j}}e^   { - {A_n}\gamma } }}{{j!\left( {l - j} \right)!}}} } } }  \Bigg\{ \left( {{A_n} - \frac{{l - j}}{\gamma }} \right)\notag \\
 &\times G_{t,3t + 1}^{3t + 1,0}\left[ {\left. {\frac{{{{\left( {dab} \right)}^t}{A_n}c}}{{{\kappa _t}{t^{2t}}}}\gamma } \right|\begin{array}{*{20}{c}}
{\omega}\\
{\tau,j}
\end{array}} \right] \notag \\
&- \frac{1}{\gamma }G_{t + 1,3t + 2}^{3t + 1,1}\left[ {\left. {\frac{{{{\left( {dab} \right)}^t}{A_n}c}}{{{\kappa _t}{t^{2t}}}}\gamma } \right|\begin{array}{*{20}{c}}
{0,\omega}\\
{\tau,1}
\end{array}} \right] \Bigg\},
\end{align}
and
\begin{align}\label{eq:pdf_total_snr_result_kappa_mu}
{f_{{\gamma _{\text{eq}}}}^{{\kappa-\mu}}}\left( \gamma  \right)&= \frac{{{\xi ^2}{t^{a + b - 2}}}}{{{{\left( {2\pi } \right)}^{t - 1}}\Gamma \left( a \right)\Gamma \left( b \right){e^{A\gamma  + \kappa \mu }}}}\notag \\
&\times \sum\limits_{i = 0}^\infty  {\sum\limits_{l = 0}^{\mu  + i - 1} {\sum\limits_{j = 0}^l {\frac{{{{\left( {A\gamma } \right)}^{l - j}}{{\left( {\kappa \mu } \right)}^i}}}{{i!j!\left( {l - j} \right)!}}} } }   \Bigg\{ \left( {{A} - \frac{{l - j}}{\gamma }} \right)\notag \\
 &\times G_{t,3t + 1}^{3t + 1,0}\left[ {\left. {\frac{{{{\left( {dab} \right)}^t}{A}c}}{{{\kappa _t}{t^{2t}}}}\gamma } \right|\begin{array}{*{20}{c}}
{\omega}\\
{\tau,j}
\end{array}} \right] \notag \\
&- \frac{1}{\gamma }G_{t + 1,3t + 2}^{3t + 1,1}\left[ {\left. {\frac{{{{\left( {dab} \right)}^t}{A}c}}{{{\kappa _t}{t^{2t}}}}\gamma } \right|\begin{array}{*{20}{c}}
{0,\omega}\\
{\tau,1}
\end{array}} \right] \Bigg\}.
\end{align}
Note that \eqref{eq:pdf_total_snr_result} and \eqref{eq:pdf_total_snr_result_kappa_mu} can be reduced to \cite[Eq. (11)]{Zedini2015performance} and \cite[Eq. (4)]{Ansari2013impact} for the cases of Nakagami-$m$/$\Gamma\Gamma$ and Rayleigh/$\Gamma\Gamma$ fading, where the IM/DD scheme is considered, respectively.

\section{Performance Analysis}\label{se:performance_analysis}
We derive novel and analytical expressions of the end-to-end outage probability and average BER for the mixed RF/FSO dual-hop system in this section. Moreover, the asymptotic outage probability in the high-SNR regime and BER performance without pointing errors are also assessed.

\subsection{Outage Probability}
As an significant metric for the performance of a wireless communication system, the outage probability is defined as the probability that the instantaneous SNR $\gamma_\text{eq}$ falls below a predetermined threshold $\gamma_\text{th}$ \cite{simon2005digital}. Mathematically, the outage probability of dual-hop AF relaying can be obtained by setting $\gamma = \gamma_\text{th}$ in \eqref{eq:cdf_total_snr_result} as ${P_{\text{o}}} = \Pr \left( {{\gamma _{\text{eq}}}} < {\gamma _{\text{th}}} \right) = {F_{{\gamma _{\text{eq}}}}}\left( {{\gamma _{\text{th}}}} \right)$.

In order to provide more insights into the impact of system and fading parameters on the system performance, we now focus on the high-SNR regime. By taking $\kappa_t \gg 1$ in \eqref{eq:cdf_total_snr_result} and \eqref{eq:cdf_total_snr_result_kappa_mu}, and using \cite[Eq. (9.303)]{gradshtein2000table}, the asymptotic high-SNR outage probability can be derived as
\begin{align}\label{eq:outage_high}
{P_{\text{o},\kappa_t \gg 1}^{{\eta-\mu}}}&= 1 - \frac{{{\xi ^2}{t^{a + b - 2}}}}{{\Gamma \left( \mu  \right)\Gamma \left( a \right)\Gamma \left( b \right){{\left( {2\pi } \right)}^{t - 1}}}}{\left( {\frac{h}{H}} \right)^\mu }\notag \\
&\times \sum\limits_{n = 1}^2 {\sum\limits_{k = 0}^{\mu  - 1} {\sum\limits_{l = 0}^{\mu  - k - 1} {\sum\limits_{j = 0}^l {\sum\limits_{i = 1}^{3t + 1} {\frac{{{a_{n,k}}{{\left( {{A_n}{\gamma _{\text{th}}} } \right)}^{l - j}}}}{{j!\left( {l - j} \right)!e^  { - {A_n}{\gamma _{\text{th}}} }  }}} } } } } \notag \\
&\times {\left( {\frac{{{{\left( {dab} \right)}^t}{A_n}c}}{{{\kappa _t}{t^{2t}}}}{\gamma _{\text{th}}} } \right)^{{\tau _i}}}\frac{{\prod\nolimits_{\rho  = 1,\rho  \ne i}^{3t + 1} {\Gamma \left( {{\tau _\rho } - {\tau _i}} \right)} }}{{\prod\nolimits_{\rho  = 1}^t {\Gamma \left( {{\omega _\rho } - {\tau _i}} \right)} }},
\end{align}
and
\begin{align}\label{eq:outage_high_kappa_mu}
{P_{\text{o},\kappa_t \gg 1}^{{\kappa-\mu}}}&= \sum\limits_{i = 0}^\infty  {\frac{{{{\left( {\kappa \mu } \right)}^i}}}{{i!{e^{k\mu }}}}}  - \frac{{{\xi ^2}{t^{a + b - 2}}}}{{{{\left( {2\pi } \right)}^{t - 1}}\Gamma \left( a \right)\Gamma \left( b \right){e^{A\gamma _{\text{th}}  + \kappa \mu }}}}\notag \\
&\times \sum\limits_{i = 0}^\infty  {\sum\limits_{l = 0}^{\mu  + i - 1} {\sum\limits_{j = 0}^l {\frac{{{{\left( {A\gamma _{\text{th}} } \right)}^{l - j}}{{\left( {\kappa \mu } \right)}^i}}}{{i!j!\left( {l - j} \right)!}}} }} \notag \\
&\times {\left( {\frac{{{{\left( {dab} \right)}^t}{A}c}}{{{\kappa _t}{t^{2t}}}}{\gamma _{\text{th}}} } \right)^{{\tau _i}}}\frac{{\prod\nolimits_{\rho  = 1,\rho  \ne i}^{3t + 1} {\Gamma \left( {{\tau _\rho } - {\tau _i}} \right)} }}{{\prod\nolimits_{\rho  = 1}^t {\Gamma \left( {{\omega _\rho } - {\tau _i}} \right)} }},
\end{align}
where ${\tau _i}$ denotes the $i$-th term of $\tau$, and ${\omega _\rho}$ accounts for the $\rho$-th term of $\omega$, respectively. Note that \eqref{eq:outage_high} and \eqref{eq:outage_high_kappa_mu} are in the form of simple elementary functions. It is clear to see from \eqref{eq:outage_high} and \eqref{eq:outage_high_kappa_mu} that the outage probability becomes smaller as the average electrical SNR $\kappa_t$ of the FSO link increases, which agrees with the corresponding findings presented in \cite{Lee2011performance,Ansari2013impact,samimi2013end,Zedini2015performance}.

\subsection{Bit Error Rate}
Note that the binary modulation schemes are widely used in experimental and practical FSO communication systems \cite{popoola2009bpsk,Karimi2009ber}. For an extensive list of different binary modulation schemes, a unified BER expression is given as \cite[Eq. (12)]{ansari2011new}
\begin{align}\label{eq:ber_formula}
{P_\text{b}} = \frac{{{q^p}}}{{2\Gamma \left( p \right)}}\int_0^\infty  {\exp \left( { - q\gamma } \right){\gamma ^{p - 1}}F\left( \gamma  \right)} d\gamma,
\end{align}
where the parameters $p$ and $q$ denote different modulation schemes. For example, by setting $p=1$ and $q=0.5$, the BER of non-coherent binary frequency shift keying (NBFSK) is obtained, while for coherent binary frequency shift keying (CBFSK), we should set $p=0.5$ and $q=0.5$ \cite{Trigui2009Performance}.

By substituting \eqref{eq:cdf_total_snr_result} and \eqref{eq:cdf_total_snr_result_kappa_mu} into \eqref{eq:ber_formula}, and using \cite[Eq. (7.813.1)]{gradshtein2000table}, we can obtain the analytical BER expressions as
\begin{align}\label{eq:ber_result}
{P_\text{b}^{\eta-\mu}}  &= \frac{1}{2} - \frac{{{\xi ^2}{t^{a + b - 2}}{q^p}}}{{2\Gamma \left( \mu  \right)\Gamma \left( a \right)\Gamma \left( b \right)\Gamma \left( p \right){{\left( {2\pi } \right)}^{t - 1}}}}{\left( {\frac{h}{H}} \right)^\mu }\notag \\
&\times \sum\limits_{n = 1}^2 {\sum\limits_{k = 0}^{\mu  - 1} {\sum\limits_{l = 0}^{\mu  - k - 1} {\sum\limits_{j = 0}^l {\frac{{{a_{n,k}}{{\left( {{A_n}} \right)}^{l - j}}{{\left( {q + {A_n}} \right)}^{j - p - l}}}}{{j!\left( {l - j} \right)!}}} } } }\notag \\
 &\times G_{t + 1,3t + 1}^{3t + 1,1}\left[ {\left. {\frac{{{{\left( {dab} \right)}^t}{A_n}c}}{{{\kappa _t}{t^{2t}}\left( {q + {A_n}} \right)}}} \right|\begin{array}{*{20}{c}}
{1 + j - p - l,\omega}\\
{\tau}
\end{array}} \right],
\end{align}
and
\begin{align}\label{eq:ber_result_kappa_mu}
{P_\text{b}^{\kappa-\mu}}  &=   \frac{1}{2}\sum\limits_{i = 0}^\infty  {\frac{{{{\left( {\kappa \mu } \right)}^i}}}{{i!{e^{k\mu }}}}}  - \frac{{{\xi ^2}{t^{a + b - 2}}{q^p}{e^{ - \kappa \mu }}}}{{2\Gamma \left( \mu  \right)\Gamma \left( a \right)\Gamma \left( b \right)\Gamma \left( p \right){{\left( {2\pi } \right)}^{t - 1}}}}\notag \\
&\times \sum\limits_{i = 0}^\infty  \sum\limits_{l = 0}^{\mu  + i - 1} {\sum\limits_{j = 0}^l {\frac{{{A^{l - j}}{{\left( {\kappa \mu } \right)}^i}{{\left( {q + A} \right)}^{j - p - l}}}}{{i!j!\left( {l - j} \right)!}}} } \notag \\
&\times G_{t + 1,3t + 1}^{3t + 1,1}\left[ {\left. {\frac{{{{\left( {dab} \right)}^t}Ac}}{{{\kappa _t}{t^{2t}}\left( {q + A} \right)}}} \right|\begin{array}{*{20}{c}}
{1 + j - p - l,\omega }\\
\tau
\end{array}} \right] .
\end{align}
Note that for the special case of mixed Rayleigh/$\Gamma\Gamma$ fading channels as investigated in \cite{Ansari2013impact}, \eqref{eq:ber_result} and \eqref{eq:ber_result_kappa_mu} can reduce to \cite[Eq. (14)]{Ansari2013impact} where the IM/DD scheme is considered.

Then, we consider the non-pointing errors case and recall that $d={\xi ^2}/({\xi ^2}+1)$. By taking $\xi \rightarrow \infty$ (larger $\xi$ means smaller pointing errors) and using the definition of the Meijer's $G$-function \cite[Eq. (9.301)]{gradshtein2000table}, the BER expressions can be converged to
\begin{align}\label{eq:ber_no_pointing_error}
{P_{\text{b},\xi \rightarrow \infty}^{\eta-\mu}}  &= \frac{1}{2} - \frac{{{t^{a + b - 2}}{q^p}}}{{2\Gamma \left( \mu  \right)\Gamma \left( a \right)\Gamma \left( b \right)\Gamma \left( p \right){{\left( {2\pi } \right)}^{t - 1}}}}{\left( {\frac{h}{H}} \right)^\mu }\notag \\
&\times \sum\limits_{n = 1}^2 {\sum\limits_{k = 0}^{\mu  - 1} {\sum\limits_{l = 0}^{\mu  - k - 1} {\sum\limits_{j = 0}^l {\frac{{{a_{n,k}}{{\left( {{A_n}} \right)}^{l - j}}{{\left( {q + {A_n}} \right)}^{j - p - l}}}}{{j!\left( {l - j} \right)!}}} } } } \notag \\
&\times G_{t - 1,2t + 1}^{2t + 1,1}\left[ {\left. {\frac{{{{\left( {ab} \right)}^t}{A_n}c}}{{{\kappa _t}{t^{2t}}\left( {q + {A_n}} \right)}}} \right|\begin{array}{*{20}{c}}
{1 + j - p - l}\\
{\tau '}
\end{array}} \right],
\end{align}
and
\begin{align}\label{eq:ber_no_pointing_error_kappa_mu}
{P_{\text{b},\xi \rightarrow \infty}^{\kappa-\mu}}   &= \frac{1}{2}\sum\limits_{i = 0}^\infty  {\frac{{{{\left( {\kappa \mu } \right)}^i}}}{{i!{e^{k\mu }}}}}  - \frac{{{\xi ^2}{t^{a + b - 2}}{q^p}{e^{ - \kappa \mu }}}}{{2\Gamma \left( \mu  \right)\Gamma \left( a \right)\Gamma \left( b \right)\Gamma \left( p \right){{\left( {2\pi } \right)}^{t - 1}}}}\notag \\
&\times \sum\limits_{i = 0}^\infty  \sum\limits_{l = 0}^{\mu  + i - 1} {\sum\limits_{j = 0}^l {\frac{{{A^{l - j}}{{\left( {\kappa \mu } \right)}^i}{{\left( {q + A} \right)}^{j - p - l}}}}{{i!j!\left( {l - j} \right)!}}} } \notag \\
&\times G_{t - 1,2t + 1}^{2t + 1,1}\left[ {\left. {\frac{{{{\left( {ab} \right)}^t}{A_n}c}}{{{\kappa _t}{t^{2t}}\left( {q + {A_n}} \right)}}} \right|\begin{array}{*{20}{c}}
{1 + j - p - l}\\
{\tau '}
\end{array}} \right],
\end{align}
where $\tau ' \triangleq \{\frac{a}{t},\frac{{a + t - 1}}{t},\frac{b}{t},\frac{{b + t - 1}}{t},j\}$. Once again, it is worthy to point out that \eqref{eq:ber_no_pointing_error} and \eqref{eq:ber_no_pointing_error_kappa_mu} reduce to \cite[Eq. (15)]{Ansari2013impact} when $t=2$ (i.e. IM/DD scheme). Note that the average BER expressions of non-binary modulation schemes can be also obtained by using the similar way presented here.

%\subsection{Capacity}

\section{Numerical Results}\label{se:numerical_results}

In this section, the analytical results of mixed RF/FSO systems over different system configuration scenarios and/or various channel fading conditions are presented and compared with Monte-Carlo simulations. Without loss of generality, we assume the heterodyne detection $t=1$, the threshold ${\gamma_\text{th}}=0$dB, the link distance $L=4000$m, the receiver aperture diameter $D=0.01$m, and a wavelength of $\lambda=1550$nm throughout this section \cite{peppas2013capacity}. For the mixed RF/FSO AF relay scheme with the fixed gain, we set $c=1$ at the relay node. In Monte-Carlo simulations, $10^6$ $\eta$-$\mu$ and $\Gamma \Gamma$ random samples are generated.

\begin{figure}[htbp]
\centering
\includegraphics[scale=0.63]{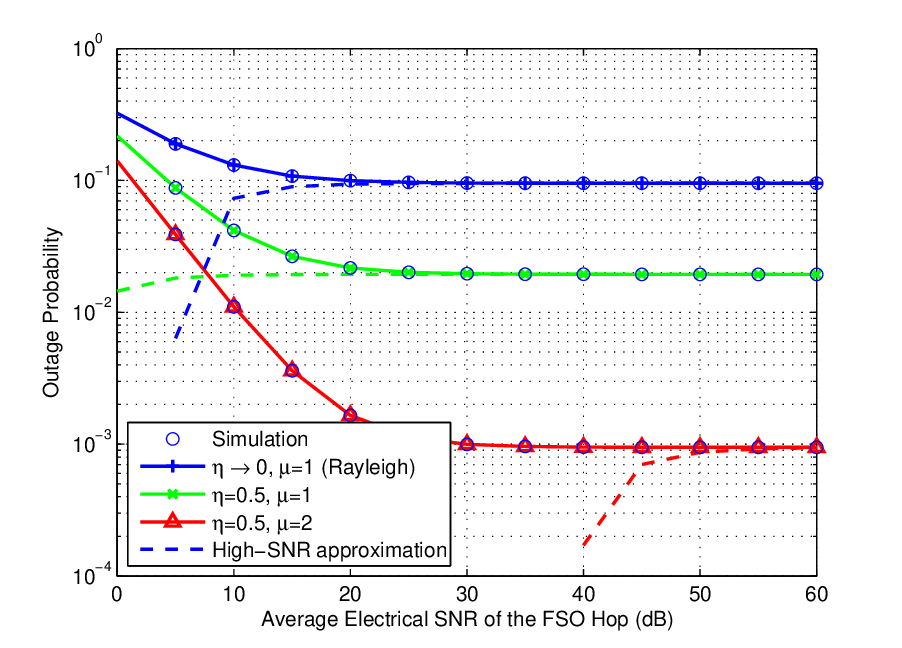}
\caption{Simulated, analytical and high-SNR approximated outage probability of mixed RF/FSO dual-hop relay systems over $\eta$-$\mu$/$\Gamma \Gamma$ fading channels versus the average electrical SNR of the FSO hop ${{\bar {\gamma_2}  }}$ ($C_n^2= 1\times 10^{-15} {\text{m}}^{-2/3}$, $D=0.01$m, $L=4000$m, $\lambda=1550$nm, $\bar {\gamma_1}=10$dB and $\xi=1.1$).
\label{fig:OP_etamu_high}}
\end{figure}

\begin{figure}[htbp]
\centering
\includegraphics[scale=0.65]{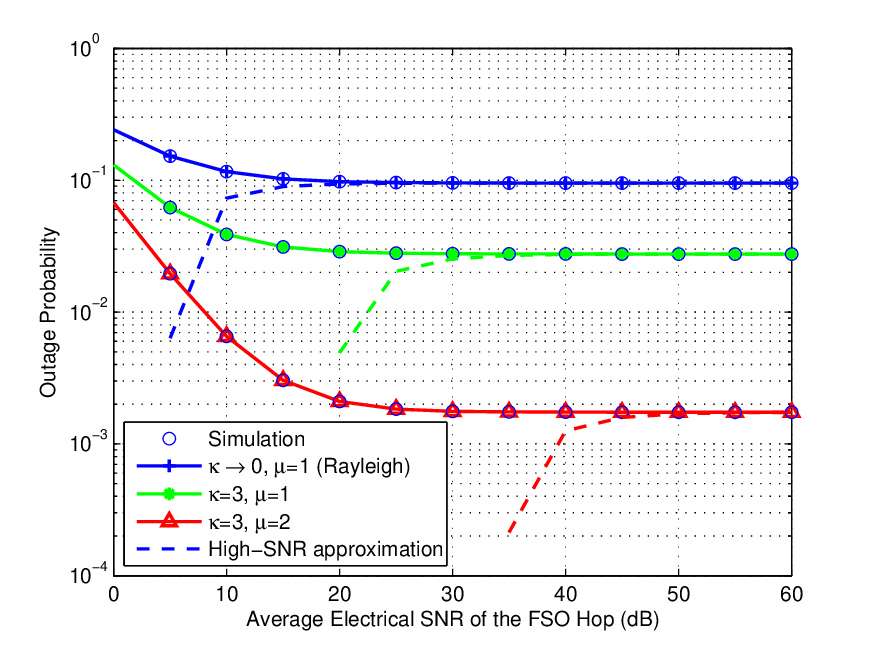}
\caption{Simulated, analytical and high-SNR approximated outage probability of mixed RF/FSO dual-hop relay systems over $\kappa$-$\mu$/$\Gamma \Gamma$ fading channels versus the average electrical SNR of the FSO hop ${{\bar {\gamma_2}  }}$ ($C_n^2= 1\times 10^{-15} {\text{m}}^{-2/3}$, $D=0.01$m, $L=4000$m, $\lambda=1550$nm, $\bar {\gamma_1}=10$dB and $\xi=1.1$).}
\label{fig:OP_kappamu}
\end{figure}

\begin{figure}[htbp]
\centering
\includegraphics[scale=0.65]{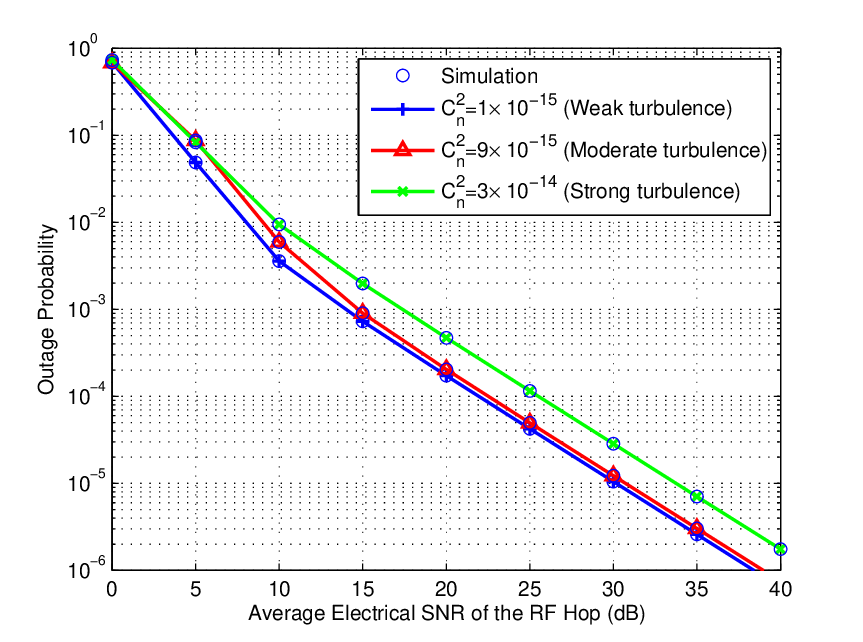}
\caption{Simulated and analytical outage probability of mixed RF/FSO dual-hop relay systems over weak to strong turbulence fading channels versus the average electrical SNR of the RF hop ${{\bar {\gamma_1}  }}$ ($\mu=3$, $\eta=0.5$, $D=0.01$m, $L=4000$m, $\lambda=1550$nm, $\bar {\gamma_2}=10$dB and $\xi=1.1$).
\label{fig:OP_Cn}}
\end{figure}

\begin{figure}[htbp]
\centering
\includegraphics[scale=0.63]{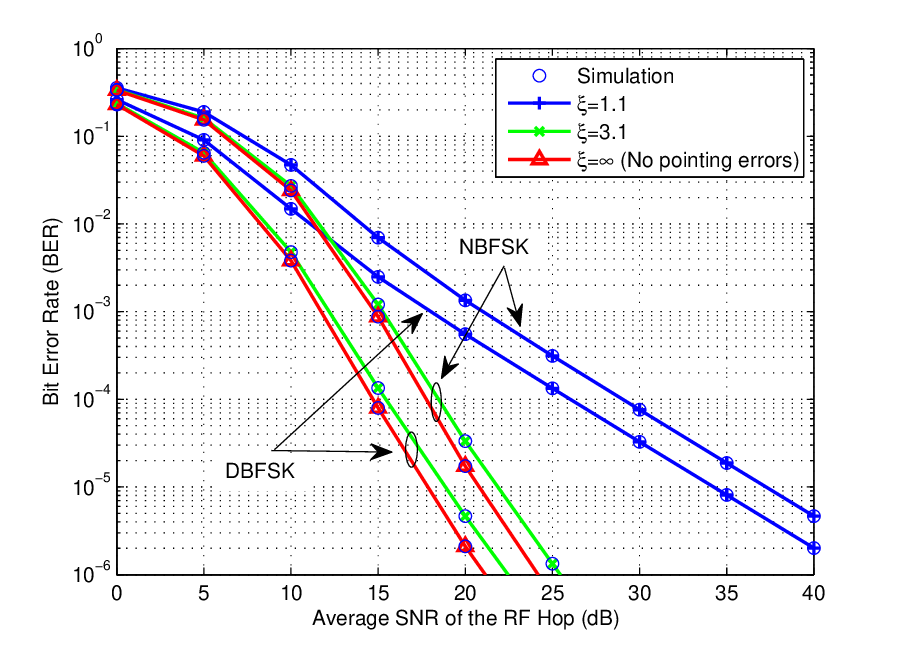}
\caption{Simulated and analytical BER of mixed RF/FSO dual-hop relay systems over $\eta$-$\mu$ and moderate turbulence fading channels versus the average electrical SNR of the RF hop ${{\bar {\gamma_1}  }}$ ($\mu=3$, $\eta=0.5$, $D=0.01$m, $L=4000$m, $\lambda=1550$nm, $\bar {\gamma_2}=10$dB and $C_n^2= 9\times 10^{-15} {\text{m}}^{-2/3}$).
\label{fig:BER_xi_DBFSK_NBFSK}}
\end{figure}

Figure \ref{fig:OP_etamu_high} depicts the simulated, analytical \eqref{eq:cdf_total_snr_result} and high-SNR approximated \eqref{eq:outage_high} outage probability against the average electrical SNR of the FSO hop when different values of the fading parameters $\mu$ and $\eta$ are considered in the RF hop. The average SNR of the RF hop is set as $\bar {\gamma_1}=10$dB, the turbulence strength is weak with $C_n^2= 1\times 10^{-15} {\text{m}}^{-2/3}$, and the pointing errors are large with $\xi=1.1$. We can find that there is a good match between the analytical and simulated results, and the high-SNR approximation is quite tight in the moderate- and high-SNR regimes, which validates the accuracy of the proposed expressions. By varying one parameter while keeping the other one fixed, Fig. \ref{fig:OP_etamu_high} reveals that increasing the values of $\mu$ and/or $\eta$ helps to overcome the fading effects. This is due to the fact that with more multipath clusters (larger $\mu$) and/or the less difference between the power of in-phase and quadrature components (larger $\eta$), the power of received signals will be enhanced. We recall that consistent conclusions were also drawn in \cite{peppas2013dual,zhang2012performance}.

It is clear to see from Fig. \ref{fig:OP_kappamu} that the simulated, analytical \eqref{eq:cdf_total_snr_result_kappa_mu} and high-SNR approximated \eqref{eq:outage_high_kappa_mu} outage probability results match well. This verifies the accuracy of the CDF expression \eqref{eq:cdf_total_snr_result_kappa_mu} derived in the analysis. With the selected parameters as shown in Fig. \ref{fig:OP_kappamu}, we find that the convergent series of \eqref{eq:cdf_total_snr_result_kappa_mu} is truncated with only ten terms to be sufficient to get numerically accurate results. In the $\kappa$-$\mu$ model, the parameter $\kappa$ indicates the power of the dominant component of a signal and the parameter $\mu$ represents the number of multipath clusters, respectively. By varying one parameter while keeping the other parameter fixed, Fig. \ref{fig:OP_kappamu} reveals that increasing the values of $\kappa$ or $\mu$ both helps overcome the effects of fading. Moreover, the effect of $\mu$ on the outage probability is more pronounced than that of $\kappa$. For example, the gap between $\kappa=3, \mu=1$ and $\kappa=3, \mu=2$ curves is larger than that between $\kappa=3, \mu=1$ and $\kappa=3, \mu=2$ curves.

In Fig. \ref{fig:OP_Cn}, the outage probability of mixed RF/FSO dual-hop relay systems under $\eta$-$\mu$/$\Gamma \Gamma$ fading conditions with $\eta=0.5$ and $\mu=3$, and strong pointing errors with $\xi=1.1$ is illustrated for various values of the strength of the atmospheric turbulence $C_n^2$. The influence of the atmospheric turbulence on the outage performance becomes more obvious from weak to strong turbulence channels. For example, at ${\bar {\gamma_1} } = 20$dB, the difference of the outage probability between $C_n^2= 9\times 10^{-15} {\text{m}}^{-2/3}$ and $C_n^2= 1\times 10^{-15} {\text{m}}^{-2/3}$ is much larger than that between $C_n^2= 3\times 10^{-14} {\text{m}}^{-2/3}$ and $C_n^2= 9\times 10^{-15} {\text{m}}^{-2/3}$. Therefore, Fig. \ref{fig:OP_Cn} indicates that the mixed RF/FSO dual-hop relay system losses much outage performance in adverse weather or strong turbulence conditions (i.e. large values of $C_n^2$).

The effect of pointing errors on the BER performance of mixed RF/FSO dual-hop relay systems with different modulation schemes is studied in Fig. \ref{fig:BER_xi_DBFSK_NBFSK}. The parameters of the RF link are $\eta=0.5$ and $\mu=3$, and the turbulence strength is $C_n^2= 9\times 10^{-15} {\text{m}}^{-2/3}$. It is clear that the analytical BER expressions coincide with the simulation results. We can also find from Fig. \ref{fig:BER_xi_DBFSK_NBFSK} that the BER of CBFSK is smaller than that of NBFSK. This observation can be explained by the fact that the system achieves better BER performance when the channel state information is available at the receivers. More importantly, the pointing errors (note that larger $\xi$ means smaller pointing errors) have an obvious effect on the BER performance, especially when $\xi$ is small. This observation agrees with the corresponding finding in \cite{Ansari2013impact}.

\section{Conclusions}\label{se:conclusion}
In this paper, we investigated a dual-hop relaying system over mixed RF/FSO links, which can be modeled as $\eta$-$\mu$/$\Gamma\Gamma$ and $\kappa$-$\mu$/$\Gamma\Gamma$ distributions. Novel and exact expressions of the CDF and PDF of the end-to-end SNR were derived in terms of finite sums of Meijer's $G$-functions. We also derive the new analytical expressions of the end-to-end outage probability and average BER for such system. These results are general, since they correspond to $\kappa$-$\mu$ and $\eta$-$\mu$ fading in the RF link, and account for pointing errors, atmospheric turbulence and different modulation schemes in the FSO link. Note that for the case of $\kappa$-$\mu$/$\Gamma\Gamma$ fading, the infinite series in corresponding results quickly and steadily converges, requiring only a few terms to obtain a desired accuracy. For example, to obtain an error smaller than $10^{-6}$, less than ten terms are required in all of considered cases. Capitalizing on the high-SNR asymptotic outage probability, some useful insights were revealed. Particularly, it is clear to find that the outage probability becomes smaller as the electrical SNR of the FSO link increases. In addition, we demonstrated that the pointing errors have an obvious effect on the BER performance.
%Finally, although the presented results are obtained for Format 1 of the $\eta$-$\mu$ fading distribution, the analytical approach can be easily extended to Format 2.

\section*{Appendix}
%\appendix[One Useful Integration Identity]
\setcounter{equation}{0}
\renewcommand{\theequation}{A.\arabic{equation}}
We involve the integral in terms of power, exponential and Meijer's $G$-functions as
\begin{align}\label{eq:integral_form}
I = \int_0^\infty  {{x^{ - \alpha  - 1}}\exp \left( { - \sigma {x^{ - 1}}} \right)} G_{p,q}^{m,n}\left[ { {\omega {x^{u/v}}} \left|\begin{array}{*{20}{c}}
{\left( {{a_p}} \right)}\\
{\left( {{b_q}} \right)}
\end{array}\right.} \right]dx,
\end{align}
where $\left( {{a_p}} \right) = \left\{ {{a_1}, \cdots ,{a_p}} \right\}$, $\left( {{b_q}} \right) = \left\{ {{b_1}, \cdots ,{b_q}} \right\}$. Applying a change of variables, $x^{-1} \rightarrow x$ and with the help of \cite[Eq. (9.31.2)]{gradshtein2000table}, \eqref{eq:integral_form} becomes
%\begin{align}\label{eq:integral_form1}
%I = \int_0^\infty  {{x^{\alpha  - 1}}\exp \left( { - \sigma x} \right)} G_{p,q}^{m,n}\left[ {\omega {x^{ - u/v}}\left| {\begin{array}{*{20}{c}}
%{\left( {{a_p}} \right)}\\
%{\left( {{b_q}} \right)}
%\end{array}} \right.} \right]dx.
%\end{align}
%
%\begin{align}\label{eq:meijer_transform}
%G_{p,q}^{m,n}\left[ {{x^{ - 1}}\left| {\begin{array}{*{20}{c}}
%{\left( {{a_p}} \right)}\\
%{\left( {{b_q}} \right)}
%\end{array}} \right.} \right] = G_{q,p}^{n,m}\left[ { x\left| {\begin{array}{*{20}{c}}
%{1 - \left( {{b_q}} \right)}\\
%{1 - \left( {{a_p}} \right)}
%\end{array}} \right.} \right],
%\end{align}
%we have
\begin{align}
I &= \int_0^\infty  {{x^{\alpha  - 1}}\exp \left( { - \sigma x} \right)} G_{q,p}^{n,m}\left[ {{\omega ^{ - 1}}{x^{u/v}}\left| {\begin{array}{*{20}{c}}
{1 - \left( {{b_q}} \right)}\\
{1 - \left( {{a_p}} \right)}
\end{array}} \right.} \right]dx \label{eq:integral_form2} \\
&= \frac{{{v^s}{u^{\alpha  - 1/2}}{\sigma ^{ - \alpha }}}}{{{{\left( {2\pi } \right)}^{\left( {u - 1} \right)/2 + {c^*}\left( {v - 1} \right)}}}}\notag \\
&\times G_{vq + u,vp}^{vn,vm + u}\left[ {\frac{{{\omega ^{ - v}}{u^u}}}{{{\sigma ^u}{v^{v\left( {p - q} \right)}}}}\left| {\begin{array}{*{20}{c}}
{\Delta \left( {u,1 - \alpha } \right),\Delta \left( {v,1 - \left( {{b_q}} \right)} \right)}\\
{\Delta \left( {v,1 - \left( {{a_p}} \right)} \right)}
\end{array}} \right.} \right], \label{eq:integral_form3}
\end{align}
where $s = \sum\limits_{j = 1}^p {\left( {1 - {a_j}} \right) - } \sum\limits_{j = 1}^q {\left( {1 - {b_j}} \right) + \frac{{q - p}}{2} + 1} $, ${c^*} = m + n - \left( {p + q} \right)/2$, and $\Delta \left( {u,v} \right) = \frac{v}{u},\frac{{v + 1}}{u}, \cdots ,\frac{{v + u - 1}}{u}$. Note that from \eqref{eq:integral_form2} to \eqref{eq:integral_form3}, we have used the identity of \cite[Eq. (2.24.3.1)]{prudnikov1990integrals3}.
By using \cite[Eq. (9.31.2)]{gradshtein2000table} again, the analytical result of \eqref{eq:integral_form} can be expressed as
\begin{align}\label{eq:integral_result}
&I= \frac{{{v^s}{u^{\alpha  - 1/2}}{\sigma ^{ - \alpha }}}}{{{{\left( {2\pi } \right)}^{\left( {u - 1} \right)/2 + {c^*}\left( {v - 1} \right)}}}}\notag \\
&\times G_{vp,vq \!+\! u}^{vm \!+\! u,vn}\left[ {\frac{{{\omega ^v}{\sigma ^u}}}{{{u^u}{v^{v\left( {q \!-\! p} \right)}}}}\left| {\begin{array}{*{20}{c}}
{1 \! -\! \Delta \left( {v,1 \!-\! \left( {{a_p}} \right)} \right)}\\
{1 \!-\! \Delta \left( {u,1 \!-\! \alpha } \right),1-\Delta \left( {v,1 \!-\! \left( {{b_q}} \right)} \right)}
\end{array}} \right.} \right].
\end{align}

%
%\bibliographystyle{IEEEtran}
%\bibliography{IEEEabrv,Ref}

%%%%%%%%%%%%%%%%%%%%%%% References %%%%%%%%%%%%%%%%%%%%%%%%%

\end{document}